\documentstyle{article}

\def\be{\begin{equation}}
\def\ee{\end{equation}}
\def\bea{\begin{eqnarray}}
\def\eea{\end{eqnarray}}
\def\al{\alpha}

\def\ep{\epsilon}
\def\vep{\varepsilon}

\def\la{\lambda}

\def\sa{\sigma}

\def\f{\frac{1}{2}}

\def\1/4{\mbox{\small $\frac{1}{4}$}}

\def\lb{\label}
\def\bb{\bibitem}
\begin{document}

\title{Variation of the speed of light due to non-minimal coupling between 
electromagnetism and gravity\footnote{Invited lecture given at the Meeting on Electromagnetism 
organized by the Fondation Louis de Broglie in Peyresq (France), September 2-7, 2002.}}
\author{P. Teyssandier \thanks{E-mail: Pierre.Teyssandier@obspm.fr}\\
\it SYRTE, CNRS/UMR 8630, Observatoire de Paris, \\  
\it 61 av. de l'Observatoire, F-75014 Paris, France}
\date{}
\maketitle

\begin{abstract}
We consider an Einstein-Maxwell action modified by the addition of three terms coupling the electromagnetic 
strength to the curvature tensor. The corresponding generalized Maxwell equations imply a variation of the 
speed of light in a vacuum. We determine this variation in Friedmann-Robertson-Walker spacetimes. 
We show that light propagates at a speed greater than $c$ when a simple condition is satisfied.

$ $

\noindent
PACS numbers: 04.40.Nr; 98.80.-k; 95.30.Sf 
\end{abstract}

\section{Introduction}

Theories with a varying speed of light have been recently proposed to solve the initial value problems of 
standard cosmology (see, e.g., \cite{Moffat1}--\cite{Moffat2} and Refs. therein). However, these theories 
break local Lorentz invariance and suffer from quite unclear physical 
interpretations. So, it seems to us that it is necessary to explore less extremist ideas.

The present paper is based on the observation that the invariant speed $c$ involved in the local Lorentz 
transformations must be carefully distinguished from the speed of light considered as a signal 
propagating in a vacuum. In fact, it is possible both to maintain the basic principles of metric 
theories of gravity and to obtain a variable speed of light by modifying Maxwell's equations \cite{Bel}. 
For this reason, we suggest to call $c$ the "spacetime structure constant" instead of the "speed of light", 
which is a very misleading terminology. 
   
In what follows, we consider the modified Einstein-Maxwell equations which arise from an electromagnetic field 
non-minimally coupled to the gravitational field. Several kinds of non-minimal couplings may be proposed 
\cite{Goenner}, some of them violating gauge invariance \cite{Novello2}. In order to avoid such a radical 
consequence, we define the action as
\begin{equation} \lb{1}
{\cal I} = \int \left[ -\frac{c^{3}}{16 \pi G} \, (R + 2\Lambda) +  L_{EM} - j^{\mu}A_{\mu} + L_{matter} \right]  
\sqrt{-g} d^{4}x  \, ,
\end{equation}
where $G$ is the Newtonian gravitational constant, $\Lambda$ is the cosmological constant, $R$ is the scalar 
curvature, $j^{\mu}$ is the current density vector, and $A_{\mu}$ is the vector potential. The non-minimally 
coupled electromagnetic field Lagrangian is defined as  
\begin{equation} \label{2}    
 L_{EM} = -\frac{1}{4} F_{\mu \nu}F^{\mu \nu} + \frac{1}{4} \xi R F_{\mu \nu}F^{\mu \nu} 
+ \f \eta R_{\mu\nu}F^{\mu\rho}F^{\nu}_{ . \rho} + \frac{1}{4} \zeta R_{\mu\nu\rho\sigma}F^{\mu\nu}
F^{\rho\sigma}  \, ,
\end{equation}
where $F_{\mu\nu} = \partial_{\mu} A_{\nu} - \partial_{\nu} A_{\mu} $ is the electromagnetic strength, 
$R_{\mu\nu\rho\sigma}$ is the curvature tensor, $R_{\mu \nu}$ is the Ricci tensor, $\xi , \eta$, and 
$\zeta$ are constants having dimensions [length]$^{2}$. The field variables are the metric $g_{\mu\nu}$, 
the vector potential $A_{\mu}$ and the variables describing matter in $L_{matter}$.

It has been previously noted that the generalized Maxwell equations deduced from 
Eqs. (\ref{1})--(\ref{2}) imply a variable speed of light \cite{Drummond1,Novello1,Lafrance}, except in the case 
where $\eta = \zeta = 0$ \cite{Novello3}. However, as far as we know, the formula giving the 
speed of light in the Friedmann-Robertson-Walker (FRW) cosmological models has not been found in the general case. 
The main purpose of the present paper is to yield this formula. A detailed analysis of its cosmological 
implications will be developed elsewhere.  

The plan of the paper is as follows. In Sect. 2, we recall why a varying speed of light is compatible 
with the general axioms underlying special and general relativity. Section 3 is devoted to arguments 
in favour of the non-minimal coupling (NMC) examined here. In Sect. 4, we derive the equations of motion 
for the electromagnetic strength $F_{\mu\nu}$ in any spacetime. In Sect. 5, we form the equations satisfied 
by the wave vector in the limit of the geometric optics approximation, with a special emphasis on the 
case where the Einstein tensor has the structure corresponding to a perfect fluid. In Sect. 6, we outline 
the theory of light rays in the FRW background. In Sect. 7, we 
get the explicit expression of the speed of light as a function of the energy and pressure
content of the Universe when Einstein equations are satisfied. Finally, we give some concluding remarks 
in Sect. 8.

{\it Conventions and notations}.-- The signature of the Lorentzian metric $g$ is $ (+ - - -)$. Greek 
indices run from 0 to 3. We put $x^0 = ct$, $t$ being a coordinate time. Given a vector or a tensor $T$, 
we often use the notations $T_{,\al} = \partial_{\al} T$ and $T_{;\al} = \nabla_{\al}T$. The curvature 
tensor $R_{\mu\nu \rho \sa}$ is defined by $w_{\mu ;\rho ;\sa} - w_{\mu ;\sa ;\rho} = - R_{\mu\nu \rho \sa} 
w^{\nu}$. For the Ricci tensor, we take $R_{\mu\nu} = R^{\la}_{ . \mu \la \nu}$. Given two 4-vectors $v$ 
and $w$, we often use the notation $(v\cdot w) = g_{\mu\nu} v^{\mu} w^{\nu}$. When there is no ambiguity, 
we write $l^{2}  = (l \cdot l) = g_{\mu\nu} l^{\mu} l^{\nu}$. We put $\kappa = 8 \pi G/c^{4}$.

\section{Variable speed of light and relativity}

Let us give some additional arguments which justify the necessity to distinguish the spacetime 
structure constant $c$ from the speed of light. Our analysis is inspired by \cite{Levy-Leblond}.
 
The widely held idea that $c$ must be identified with the speed of light in relativistic theories is supported 
by the fact that the statement of invariance of the speed of light played a crucial role in the original 
Einstein's paper \cite{Einstein}. However, this earliest approach is not the more logical one and 
can be criticized for several reasons. 

1) If we state that the invariant quantity $c$ is linked to a property of electromagnetic radiation, 
it is hard to understand why all interactions are governed, at least locally, by special 
relativity.

2) The invariance of the laws of physics under the Poincar\'e group allows the existence of zero-mass 
particles but does not imply that any empirical interaction must be mediated by a zero-mass particle. As 
a consequence, if once a day a non-zero mass is found for the photon, only the usual presentations of special 
relativity based on the invariance of the speed of light will be disproved.  
     
3) Shortly after the Einstein's paper, it was pointed out in \cite{Ignatowski,Frank} 
that it was not necessary to assume the constancy of the light velocity in order to derive the Lorentz 
transformations. Since these pioneering works, several other derivations of the correct transformation 
equations have been performed without imposing the existence of an invariant speed 
(see, e.g., \cite{Berzi} and Refs. therein). In \cite{Levy-Leblond}, it 
was proved by elementary considerations that if {\it i)} the principle of relativity is valid, 
{\it ii)} spacetime is a four-dimensional manifold, {\it iii)} spacetime is homogeneous, {\it iv)} space 
is isotropic, {\it v)} the inertial transformations constitute a group and {\it vi)} causality is 
preserved, then the tranformation equations may be written as follows in the 
two-dimensional case:
\begin{equation} \label{3}
x' = \frac{x - vt}{\sqrt{1 -\chi v^2}} \, , \quad \quad  t' = \frac{t - \chi vx}{\sqrt{1 -\chi v^2}} \, ,
\end{equation}
where $\chi$ is a constant which must be $\geq 0$ ($\chi<0$ would imply violations of causality). 
The Galileo transformations are recovered for $\chi = 0$, while the Lorentz transformations are 
obtained for $\chi >0$. Clearly, Eq. (\ref{3}) implies that the quantity 
$c = 1/\sqrt{\chi}$ is both an invariant and a limiting speed.

The above-mentioned axioms are universal principles which do not require any definite, well-developed 
theory of some physical interaction. The existence of such axioms proves that the constant $c$ 
appearing in the Lorentz transformations is not essentially the speed of light in a vacuum, but is in fact 
a structural constant which characterizes the four-dimensional continuum constituting the arena of physical 
events. 

It follows from this discussion that a variable speed of light cannot be forbidden by special or general 
relativity. 

\section{Arguments in favour of non-minimal coupling}

Several arguments may be given in favour of the non-minimal coupling defined by Eqs. (\ref{1})-(\ref{2}).

1) It may be argued from a theorem due to Horndeski \cite{Horndeski1, Horndeski2} that the most general 
electromagnetic equations which are {\it i)} derivable from a variational principle, {\it ii)} at most of 
second-order in the derivatives of both $g_{\mu\nu}$ and $A_{\mu}$, {\it iii)} consistent with the charge 
conservation, and {\it iv)} compatible with Maxwell's equations in flat 
space-time are given by a Lagrangian defined by Eq. (2) with
\begin{equation} \label{4}
\eta =  -2 \xi \, , \quad   \zeta = \xi  \, ,
\end{equation}  
$\xi$ being arbitrary. This beautiful theorem proves that a particular NMC is inevitable if one considers the 
generalization of Maxwell's equations compatible with the currently accepted principles of electromagnetism. 
The interest of the Horndeski Lagrangian is enhanced by the fact that it can be recovered from the Gauss-Bonnet 
action in a five-dimensional space with a Kaluza-Klein metric (see \cite{Buchdahl}, \cite{Muller-Hoissen} and 
Refs. therein). 

2) It may also be argued that couplings to the curvature are induced by vacuum polarization in quantum 
electrodynamics(QED) \cite{Drummond1, Daniels}. Indeed, vacuum polarization confers a size to the photon 
of the order of the Compton wavelength of the electron. So the motion of the photon is influenced by a tidal 
gravitational effect depending on the curvature. Working in the one-loop approximation, Drummond and Hathrell 
found an effective Lagrangian for QED given by Eq. (\ref{2}) with 
\begin{equation} \label{5}
\xi = -\frac{\al}{36\pi} \lambda_{c} \! \! \! \! \bar{} \; {}^{2}  \, ,
\quad \eta = \frac{13\al}{180\pi} \lambda_{c} \! \! \! \! \bar{} \; {}^{2}  \, , 
\quad \zeta = - \frac{\al}{90\pi} \lambda_{c} \! \! \! \! \bar{} \; {}^{2}  \, ,     
\end{equation}
where $\alpha$ is the fine-structure constant and $\lambda_{c} \! \! \! \! \bar{} \, \, \, \,$ is the Compton 
wavelength of the electron defined as $\lambda_{c} \! \! \! \! \bar{} \, \, \, = \hbar /m_e c$ \cite{Note1}. 

We think that these remarkable results show that the NMC introduced by Eqs. (\ref{1})--(\ref{2}) deserves 
to be studied in detail.


\section{Equations of generalized electromagnetism}


Varying the action ${\cal I}$ with respect to the vector potential $A_{\mu}$ leads to generalized Maxwell 
equations. It is easy to form these equations by using the following lemma. Let $L$ be a scalar 
Lagrangian such that
\begin{equation} \label{6}
L = \frac{1}{4} E^{\mu\nu\rho\sa} F_{\mu\nu} F_{\rho\sa}- j^{\mu}A_{\mu} \equiv
E^{\mu\nu\rho\sa}A_{\mu ;\nu}A_{\rho ;\sa} - j^{\mu}A_{\mu}  \, ,
\end{equation}
where $E^{\mu\nu\rho\sa}$ is a 4-rank tensor which involves neither the vector potential $A_{\mu}$ nor its 
derivatives of any order and which satisfies the properties of symmetry and antisymmetry
\begin{equation} \label{7}
E^{\mu\nu\rho\sa} = E^{\rho\sa\mu\nu} \, ,
\end{equation}
\begin{equation} \label{8}
E^{\mu\nu\rho\sa} = - E^{\nu\mu\rho\sa} = - E^{\mu\nu\sa\rho} \, .
\end{equation} 
It is easily seen that the corresponding Euler-Lagrange equations may be written as 
\begin{equation} \label{9}
\nabla_{\nu}(E^{\mu\nu\rho\sa}F_{\rho\sa}) = j^{\mu} \, .
\end{equation}

Applying this lemma to the Lagrangian $L_{EM}$ defined by Eq. (\ref{2}) and using Eq. (\ref{9}), we get the 
equations of motion
\begin{equation} \label{10}
\widetilde{F}^{\mu\nu}_{\quad ; \nu} = - j^{\mu} \, ,
\end{equation}
where $ \widetilde{F}^{\mu\nu}$ is the 2-rank tensor   
\begin{equation} \label{11}
\widetilde{F}^{\mu\nu} = \left(1- \xi R \right) F^{\mu\nu} - \eta 
\left( R_{\la}^{\mu}F^{\la \nu} - R_{\la}^{\nu}F^{\la \mu} \right) - \zeta R^{\mu\nu\rho\sigma}F_{\rho\sigma} \, .
\end{equation}

Of course, Eq. (\ref{10}) must be complemented by the equations 
\begin{equation} \label{12}
F_{\mu\nu ; \rho} +F_{\rho\mu ; \nu} 
+ F_{\nu\rho ; \mu} = 0 \, .
\end{equation}
 
Since the tensor $\widetilde{F}^{\mu\nu}$ defined by Eq. (\ref{11}) is obviously antisymmetric, the charge 
conservation equation $\nabla_{\al} j^{\al} = 0$ is a condition of integrability of Eqs. (\ref{11})-(\ref{12}). 
So, charge conservation is embodied in the NMC theory deduced from the action ${\cal I}$. 

It follows from Eq. (\ref{10}) that $\widetilde{F}^{\mu\nu}$ may be 
considered as the electromagnetic excitation \cite{Hehl}. Thus, the non-minimally coupled electromagnetic 
excitation in a vacuum must be distinguished from the electromagnetic strength $F^{\mu\nu}$. Using the 
identities
\[  
R^{\mu\nu\rho\sa}_{\quad \, \, \, ; \nu}F_{\rho\sa} \equiv 2R^{\mu}_{\rho ; \sa} F^{\rho\sa} \, ,
\]
it is easily seen that Eq. (\ref{10}) may be written as
\bea 
& & (1 - \xi R)F^{\mu\nu}_{\quad  ; \nu} - \eta (R^{\mu}_{\la}F^{\la\nu}_{\quad  ; \nu} - 
R^{\nu}_{\la}F^{\la\mu}_{\quad  ; \nu})  - \zeta R^{\mu\nu}_{. \, . \, \rho \sa}F^{\rho\sa}_{\quad  ; \nu}  
\nonumber \\
& & \qquad \qquad  \qquad   - \f (2 \xi + \eta)R_{,\nu} F^{\mu\nu} - 
(\eta + 2 \zeta) R^{\mu}_{\rho ; \sa} F^{\rho\sa} = - j^{\mu} \, .  \label{13}
\eea

It is worthy of note that these equations involve the third partial derivatives of the metric, except of 
course in the case where Eqs. (\ref{4}) are satisfied. 

Introducing now the Weyl tensor $ C^{\mu\nu\rho\sigma}$ defined by
\bea  
& &C^{\mu\nu\rho\sigma} = R^{\mu\nu\rho\sigma} - \frac{1}{2} \left(R^{\mu\rho} g^{\nu\sa} + 
R^{\nu\sigma} g^{\mu\rho} - R^{\mu\sa} g^{\nu\rho} - R^{\nu\rho} g^{\mu\sa} \right)  \nonumber \\
&  & \qquad  \qquad \qquad  \qquad \qquad +\frac{1}{6}R \left( g^{\mu\rho} g^{\nu\sa} -  g^{\mu\sa} 
g^{\nu\rho} \right) \, , \label{14}
\eea
Eq. (\ref{13}) becomes
\bea \label{15}
& &\left[1 - \frac{1}{3} (3\xi - \zeta) R \right] F^{\mu\nu}_{\quad  ; \nu} 
- (\eta + \zeta)(R^{\mu}_{\la}F^{\la\nu}_{\quad  ; \nu} - 
R^{\nu}_{\la}F^{\la\mu}_{\quad  ; \nu}) - \zeta C^{\mu\nu}_{ . \, . \, \rho\sa} F^{\rho\sa}_{\quad  ; \nu} 
\nonumber \\
& & \qquad \qquad  \qquad   - \f (2 \xi + \eta)R_{,\nu} F^{\mu\nu} - 
(\eta + 2 \zeta) R^{\mu}_{\rho ; \sa} F^{\rho\sa} = - j^{\mu} \, . 
\eea

This last form of the equations of motion will be very useful to study the propagation of light in a 
gravitational wave or in the FRW background.


\section{Geometric optics approximation}


In order to determine the speed of light, we shall work in the limit of the geometric optics 
approximation. Treating the vector potential $A_{\mu}$ as the real part of a complex vector, we suppose that 
there exist solutions to Eq. (\ref{15}) which admit a development of the form
\begin{equation} \label{16}
A_{\mu}(x, \vep) = \Re \left\{ [a_{\mu}(x) + O(\ep)]\exp \left(\frac{i}{\vep}\widehat{S}(x) \right) \right\}  
\, , 
\end{equation}
where $a_{\mu}$ is a slowly varying, complex vector amplitude, $\widehat{S}(x)$ is a real function and $\vep$ is a 
dimensionless parameter which tends to zero as the typical wavelength of the electromagnetic signal becomes 
shorter and shorter. Let us define the wave vector $l_{\mu}$ as 
\begin{equation} \label{17}
l_{\mu} = \frac{1}{\vep}\widehat{S}_{, \mu} \,  .
\end{equation}
We have
\begin{equation} \label{17a}
F_{\mu\nu} = \Re \left\{i(l_{\mu} a_{\nu} - l_{\nu} a_{\mu})\exp 
\left(\frac{i}{\vep}\widehat{S}(x) \right)   \right\} + \cdots  \, .
\end{equation}
Inserting Eq. (\ref{17a}) into Eq. (\ref{15}), and then retaining only the leading terms of order $\vep^{-2}$, 
we obtain the equations constraining the wave vector $l^{\mu}$ in the form:
\bea \label{18}
& & \left\{ \left[1 - \left(\xi - \frac{1}{3}\zeta \right)R \right]l^2 - (\eta + \zeta)Ric(l, l) \right\}a^{\mu}
- (\eta + \zeta) R_{\la}^{\mu}\, l^2 \, a^{\la}  \nonumber \\
& & \qquad  \, \, - \left\{ \left[1 - \left(\xi - \frac{1}{3}\zeta \right)R \right](a\cdot l) 
- (\eta + \zeta)Ric(a, l)  \right\} l^{\mu}  \nonumber \\
& & \qquad \qquad \quad \quad + (\eta + \zeta)R_{\la}^{\mu}\, (a \cdot l)\, l^{\la}
 - 2 \zeta C^{\mu\nu\rho\sa}l_{\nu}a_{\rho}l_{\sa} = 0 \, ,
\eea
where we use the notation $Ric(v, w) = R_{\mu\nu}v^{\mu}w^{\nu}$. 

Equation (\ref{18}) shows that the wave vector is generally not a null vector and that $l^2$ will depend on the 
polarization vector $ f^{\mu} = a^{\mu} / a$, $a$ being the scalar amplitude defined by 
$a = \sqrt{\mid \! a^{\mu}\bar{a}_{\mu} \! \mid }$. Thus, light rays are not null geodesics and a gravitational 
field has properties of birefringence \cite{Drummond1,Lafrance,Balakin}. Moreover, Eq. (\ref{18}) remain 
invariant under scaling of the wave vector $l_{\mu}$. As a consequence, the photon trajectories are 
frequency independent: the gravitational field is not dispersive.

Let us now restrict our attention to the case where the Ricci tensor is of the form corresponding 
to a perfect fluid in general relativity. This means that there exists a unit timelike vector $u^{\mu}$ 
such that $R_{\mu\nu}$ may be written in the form
\begin{equation} \label{19}
R_{\mu\nu} = \frac{1}{3}(4U - R)u_{\mu} u_{\nu} - \frac{1}{3}(U-R)g_{\mu\nu} \, ,
\end{equation}
where $U$ is a scalar function. This scalar function is such that 
\begin{equation} \label{20}
U = Ric(u, u) \, .
\end{equation}
Then, defining ${\cal A}$ and ${\cal B}$ as
\begin{equation} \label{21}
{\cal A} = - \frac{1}{3}(3\xi + 2\eta + \zeta)R + \frac{2}{3}(\eta + \zeta)U \, , \quad 
{\cal B} =  \frac{1}{3}(\eta + \zeta)(4U - R) \, ,
\end{equation}
Equation (\ref{18}) reduces to 
\bea \label{22}
& & \left[(1 + {\cal A})l^2 - {\cal B}(u \cdot l)^2 \right] a^{\mu}
- \left[(1 + {\cal A})(a \cdot l) - {\cal B}(u \cdot a)(u \cdot l) \right] l^{\mu} \nonumber \\
& &  \quad - {\cal B}\left[ (u \cdot a)l^2 - (u \cdot l)(a \cdot l) \right] u^{\mu} - 
2 \zeta C^{\mu\nu\rho\sa}l_{\nu}a_{\rho}l_{\sa} = 0 \, . 
\eea
Contracting Eq. (\ref{22}) by $u_{\mu}$ yields the relation
\begin{equation} \label{23}
(1 + {\cal A} - {\cal B})\left[(u\cdot a)l^2 - (a\cdot l)(u\cdot l)\right] = 2\zeta C(u, l, a, l)\, ,
\end{equation}
where
\begin{equation} \label{23a}
C(u, l, a, l) = C^{\mu\nu\rho\sa} u_{\mu}l_{\nu}a_{\rho}l_{\sa} \, .
\end{equation}
Eliminating $(u\cdot a)l^2 - (a\cdot l)(u\cdot l)$ between Eq. (\ref{22}) and Eq. (\ref{23}), we find  
\bea \label{24}   
& & \left[(1 + {\cal A})l^2 - {\cal B}(u \cdot l)^2 \right] a^{\mu}
- \left[(1 + {\cal A})(a \cdot l) - {\cal B}(u \cdot a)(u \cdot l) \right] l^{\mu} \nonumber \\
& &  \qquad - 2\zeta \frac{{\cal B}}{1 + {\cal A} - {\cal B}}\, C(u, l, a, l) u^{\mu} 
- 2 \zeta C^{\mu\nu\rho\sa}l_{\nu}a_{\rho}l_{\sa} = 0 \, .
\eea
It is easily seen that Eq. (\ref{24}) are equivalent to Eq. (\ref{22}) if the inequalities 
\begin{equation} \label{25} 
1 + {\cal A}- {\cal B} \neq 0 \, , \quad  1 + {\cal A} \neq 0 \, 
\end{equation}
are satisfied.


\section{Application to FRW cosmological models}


In what follows, we assume that the field $F_{\mu\nu}$ is a test field propagating in a 
Friedmann-Robertson-Walker (FRW) universe with a metric 
\begin{equation} \label{26}
ds^2 = (dx^0)^{2} - a^{2}(x^{0})\left[ \frac{dr^{2}}{1-kr^{2}} + r^{2}(d\theta^{2} 
+ \sin^{2}\theta d\varphi ^{2}) \right] \, , 
\end{equation}
where $a(x^{0})$ is the scale factor and $k= 0, 1, -1$ for flat, closed and open models, 
respectively. In these models, the Ricci tensor may be written in the form given 
by Eq. (\ref{19}), where $u^{\mu}$ is the unit 4-velocity of a comoving observer (observer 
moving with the average flow of cosmic energy). Moreover, the metric $g$ is conformally flat, 
which means that
\begin{equation} \label{27}
C^{\mu\nu\rho\sigma} = 0 
\end{equation}  
throughout spacetime. As a consequence, Eq. (\ref{24}) reduces to
\begin{equation} \label{28}
\left[(1 + {\cal A})l^2 - {\cal B}(u \cdot l)^2 \right] a^{\mu}
- \left[(1 + {\cal A})(a \cdot l) - {\cal B}(u \cdot a)(u \cdot l) \right] l^{\mu} = 0 \, .
\end{equation}
Of course, we suppose that inequalities (\ref{25}) hold.
 
If it is assumed that $(1+{\cal A})l^2 - {\cal B}(u\cdot l)^2 \neq 0$, Eq. (\ref{28}) yields
\[
a^{\mu} = \frac{(1 + {\cal A})(a \cdot l) - {\cal B}(u \cdot a)(u \cdot l)}
{(1 + {\cal A})l^2 - {\cal B}(u \cdot l)^2 }l^{\mu}  \, ,
\]
which implies $F_{\mu\nu} = 0$ (see Eq. (\ref{17a})). As a consequence, the wave vector must fulfill 
the condition
\begin{equation} \label{31}
(1+{\cal A})l^2 - {\cal B}(u\cdot l)^2 = 0 \, .
\end{equation}

Two cases are to be envisaged.

1. {\it General case:} ${\cal B} \neq 0 \, $.-- It results from Eq. (\ref{31}) that $l^2 \neq 0$ and 
$(u\cdot l)\neq 0$ (indeed, $l^2 = 0$ would imply $(u\cdot l) = 0$, which is impossible if $l \neq 0$). 
Therefore, the phase velocity of light is not equal to the fundamental constant $c$ if $\eta + \zeta \neq 0$ 
and $4U - R \neq 0$. Since $l^2 \neq 0$, it is possible to choose the gauge so that the Lorentz condition
\begin{equation} \label{32}
(a \cdot l) = 0 
\end{equation}
is satisfied. With this choice, Eq. (\ref{28}) gives 
\begin{equation} \label{33}
(u\cdot a) = 0 \, .
\end{equation}
The corresponding polarization vector is orthogonal to the unit 4-velocity $u^{\mu}$ and to the wave vector 
$l_{\mu}$.

2. {\it Case where} ${\cal B} = 0 \,$.-- The wave vector $l$ is then a null vector, as in the usual 
geometric optics approximation. It is worthy of note that Eq. (\ref{32}) is now a consequence of 
Eq. (\ref{28}) and does not result from a special choice of gauge. Nevertheless, the gauge may be 
chosen so that Eq. (\ref{33}) is satisfied.

It follows from Eq. (\ref{31}) that the phase speed of light with respect to a comoving observer has the same 
value $c_{l}$ in all directions and for all polarizations. Since the gravitational field is not 
dispersive, $c_{l}$ is also the group speed of light. So we shall simply call 
$c_{l}$ the speed of light with respect to a comoving observer without any further specification. Using the 
general theory of geometric optics 
exposed in \cite{Synge}, we deduce from Eq. (\ref{31}) that the ratio $c_{l}/c $ is determined by
\begin{equation} \label{34}
\frac{c_{l}^{2}}{c^{2}} = \frac{1 + {\cal A}}{1 + {\cal A} - {\cal B}} = 
1 + \frac{(\eta + \zeta) (4U - R)}{3 - (3 \xi + \eta)R - 2(\eta + \zeta)U}  \, ,
\end{equation}
where ${\cal A}$ and ${\cal B}$ are defined by Eq. (\ref{21}). This formula shows that the speed of light 
in a FRW background generically differs from $c$ except in the case where $\eta + \zeta = 0$. 

We can suppose that at present time $1+{\cal A} \approx 1$ and $1 + {\cal A} - {\cal B} \approx 1$. So, we 
shall henceforth restrict our attention to the part of the history of the Universe such that the two 
conditions 
\begin{equation} \label{34a}
1+{\cal A} > 0 \, , \quad  1 + {\cal A} - {\cal B} >0
\end{equation}
are satisfied. Indeed, it is clear that the violation of at least one of these conditions can only occur 
in a domain of spacetime where the curvature is so great that the present theory is probably no longer 
realistic. Inequalities (\ref{34a}) are sufficient to ensure that the quantity $c_{l}$ determined by Eq. 
(\ref{34}) is real. Moreover, they imply the following equivalence:
\begin{equation} \label{34b}
c_{l} >c \, \,  \Longleftrightarrow \, \, (\eta + \zeta)(4U - R) >0 \, .
\end{equation}

Equation (\ref{34}) shows that the vacuum acts as a medium moving with the unit 4-velocity $u^{\mu}$ 
and having a refractive index $n$ given by 
\begin{equation} \label{35}
n = \frac{c}{c_{l}} = \sqrt{1 - \frac{{\cal B}}{1 + {\cal A}}} \, .
\end{equation}
Using a well-known result of the geometric optics approximation \cite{Synge}, we can state that 
the light rays are null geodesics with respect to the associated metric tensor $\overline{g}$ defined by 
\begin{equation} \label{36}
\overline{g}_{\mu\nu} =  g_{\mu\nu} - \left( 1 - \frac{1}{n^{2}} \right) u_{\mu} u_{\nu}
= g_{\mu\nu} + \frac{{\cal B}}{1 + {\cal A} - {\cal B}} u_{\mu} u_{\nu} \, .
\end{equation}
However, taking into account Eqs. (\ref{26}) and (\ref{35}), it is easily seen that the conformal 
metric $d\tilde{s}^{2} = n^{2} d\overline{s}^{2}$ is a FRW metric with the scale factor 
$ \tilde{a} = n \, a$. So, we can enunciate the following theorem:

{\bf Theorem 1}.-- {\em In a FRW background, the light rays are null geodesics of the new FRW metric defined 
as} 
\begin{equation} \label{37}
d \tilde{s}^{2} = (dx^{0})^{2} - \tilde{a}^{2}(x^{0}) \left[ \frac{dr^{2}}{1-kr^{2}} 
+ r^{2}(d\theta^{2} + \sin^{2}\theta d\varphi ^{2})   \right] \, ,
\end{equation}
{\em where}
\begin{equation} \label{38}
\tilde{a}(x^{0}) = \frac{c}{c_{l}} a(x^{0}) = \sqrt{1 - \frac{{\cal B}}{1 + {\cal A}}} \, a(x^{0}) \, ,
\end{equation}
{\em with}
\begin{equation} \label{38a}
{\cal A} = 2(3 \xi + \eta)\frac{\ddot{a}}{a} + 2(3 \xi +2\eta +\zeta)
\left( \frac{\dot{a}^{2}}{a^{2}} + \frac{k}{a^{2}}\right) \, ,
\end{equation}
\begin{equation} \label{38b}
{\cal B} = 2(\eta + \zeta) \left( - \frac{\ddot{a}}{a} + \frac{\dot{a}^{2}}{a^{2}}
+ \frac{k}{a^{2}}\right) \, ,
\end{equation}
$\dot{a}$ {\em denoting the derivative} $da/dx^{0}$.

It follows from this theorem that all geometrical properties of light rays can be obtained by 
substituting $\tilde{a} (x^{0})$ for $a(x^{0})$ into the usual definitions and relations (luminosity 
distances, number counts, gravitational lensing,...). In particular, the red-shift of an extragalactic 
comoving object will be given by
\begin{equation} \label{41}
1+z = \frac{a_{0}}{a_{e}} \frac{n_{0}}{n_{e}} = \frac{a_{0}}{a_{e}} \frac{(c_{l})_{e}}{(c_{l})_{0}} \, ,
\end{equation}
where the subscripts $e$ and $0$ stand for the time of emission and for the present time, respectively. As a 
consequence, a non-minimal coupling between the electromagnetic field and gravity is able to affect our 
informations concerning the evolution of the Universe.


\section{Speed of light and energy content of FRW models}


In order to connect the speed of light with the energy content of the Universe, we now suppose that 
the metric satisfies Einstein equations. In the FRW models, the r.h.s. of Einstein equations 
may always be considered as the energy-momentum tensor $T_{\mu\nu}$ of a perfect fluid having an energy 
density $\mu$ and a pressure $p$, $\mu$ and $p$ only depending on the cosmic time. So, we have
\begin{equation} \label{42}   
R_{\mu\nu } = \kappa \left[ (\mu  + p)u_{\mu}u_{\nu} - \frac{1}{2}(\mu - p) g_{\mu\nu} \right] 
- \Lambda g_{\mu\nu} \, .
\end{equation}
As a consequence, $R$ and $U$ are given by 
\begin{equation} \label{43}
R = - \kappa (\mu  - 3p) - 4 \Lambda \, , \quad U = \frac{1}{2} \kappa (\mu  + 3p) - \Lambda  \, .
\end{equation}
Substituting for $R$ and $U$ from Eq. (\ref{43}) into Eq. (\ref{34}), we obtain the second theorem of this 
paper.

{\bf Theorem 2}.-- {\em Given a FRW model, let} $\mu$, $p$ {\em and} $c_{l}$ {\em respectively denote the 
total energy density, the pressure and the speed of light with respect to a comoving observer. Then}

-- {\em if} $ \eta + \zeta = 0$ ,
\begin{equation} \label{43a}
c_{l} = c \, ;
\end{equation}   

-- {\em if} $ \eta + \zeta \neq 0$ ,  
\begin{equation} \label{44}
c_l =  c \,  \sqrt{1 + \frac{\mu  + p}
{\mu_{m}  + \frac{1}{3} (\sigma - 2)\mu - \sigma p }}  \, ,
\end{equation}
{\em where} $\sigma$ {\em is defined as}
\begin{equation} \label{45}
\sigma = \frac{3\xi + 2 \eta + \zeta}{\eta + \zeta }
\end{equation}
{\em and} 
\begin{equation} \label{46}
\mu_{m} = \frac{1}{(\eta + \zeta)\kappa } + \frac{2}{3}(2\sigma - 1)\frac{\Lambda}{\kappa} \, .
\end{equation}

From Eq. (\ref{43}), it is easily seen that Eq. (\ref{34b}) may be written as
\begin{equation} \label{46a}
c_{l} > c \, \Longleftrightarrow  \,  (\eta + \zeta) (\mu + p) >0  \, .
\end{equation}
As a consequence, the speed of light is greater than $c$ for any reasonable equation of state if and only 
if the condition $ \eta + \zeta >0 $ is satisfied.

Neglecting the pressure, Eq. (\ref{44}) yields   
\begin{equation} \label{47}  
\frac{c_l}{c} = 1 + \frac{\mu}{2\mu_{m}} + O(\frac{\mu^{2}}{\mu_{m}^{2}}) \, .
\end{equation}
Thus, up to the first order in $\mu / \mu_{m}$, the variation of the speed of light when $p$ is 
negligible is entirely governed by the value of $\mu_{m}$. 

Using the values of $\xi$, $\eta$ and $\zeta$ given by Eq. (\ref{5}), it may be seen that 
$\eta + \zeta > 0$. So, the speed of light obtained from the Drummond-Hathrell Lagrangian is greater 
than $c$ as long as $\mu + p >0$. Neglecting the contribution of the 
cosmological constant in Eq. (\ref{46}), we find $\mu_{m}/c^{2} = 2.5 \times 10^{51} $ g.cm$^{-3}$. 
Then, taking $\mu_{0}/c^{2} \approx 2.5\times 10^{-30}$ g.cm$^{-3}$ and neglecting the pressure, we see 
that at the present time
\begin{equation} \label{48} 
\left( \frac{c_{l}}{c}\right)_{0} - 1 \,  \approx  \, 5 \times 10 ^{-82}. 
\end{equation}
As a consequence, the difference between $c_{l}$ and $c$ predicted in the one-loop approximation of QED 
cannot be detected by local experiments. 

Finally, let us note that the speed of light $c_{l} = c$ in a de Sitter spacetime whatever the 
parameters $\xi , \eta$, and $\zeta$, since one has $\mu + p=0 $ in this case.

\section{Conclusion}


In this paper, we have outlined the theory of light rays propagating in a FRW background according to 
the NMC between electromagnetism and gravity defined by Eqs. (\ref{1})--(\ref{2}). We have obtained the 
general expression of the speed of light $c_{l}$ as a function 
of the energy density and of the pressure of the Universe. We have found that light propagates at a speed 
greater than $c$ if and only if $\eta + \zeta >0$, provided that $\mu + p >0$. This conclusion generalizes a 
result previously obtained by Drummond and Hathrell in the framework of QED. 
  
An application of these results to the horizon problem in cosmology is in preparation.


\section*{Acknowledgments}   

We are deeply grateful to B. Linet for indicating to us the existence of the Horndeski's theorem and 
making several useful remarks.

\end{document}